\documentclass[twoside,fleqn]{article}
\usepackage{graphicx}
\usepackage{amsmath}
\usepackage{espcrc2}

\newcommand{\beq}{\begin{eqnarray}}
\newcommand{\eeq}{\end{eqnarray}}

\newcommand{\pardis}{\langle \mu \rangle}

\title{Chiral transition and deconfinement in $N_f = 2$ QCD}

\author{
    M. D'Elia\address[GENO]{Dipartimento di Fisica dell'Universit{\`a} di
    Genova and INFN, Via Dodecaneso 33, I-16146, Genova, Italy},
    A. Di Giacomo\address[PISA]{Dipartimento di Fisica dell'Universit\`a di Pisa    and INFN, Largo Pontecorvo 2, I-56127 Pisa, Italy}, 
    B. Lucini\address[ZUR]{Institute for Theoretical Physics, ETH Z\"urich, CH-8093 Z\"urich, Switzerland},
    G. Paffuti\addressmark[PISA], C. Pica\addressmark[PISA]} 

\begin{document}

\begin{abstract}
The transition is studied by means of a disorder parameter detecting
condensation of magnetic monopoles in the va\-cu\-um. The deconfining
transition is found to coincide with the chiral transition and
the susceptibility $\rho$, related to the disorder parameter, is
consistent with a first order phase transition.
\end{abstract}

\maketitle

\section{Introduction}
\label{sec:introduction}

Understanding confinement and deconfinement in the presence of dynamical
fermions is still an open issue in strong interaction physics. 
There is general agreement on the order-disorder nature of the deconfining
transition in the quenched case. The popular order parameter is the Polyakov 
line $\langle L \rangle$; the symmetry
involved is $Z_N$. Alternatively the dual ('t Hooft) line 
$\langle \tilde{L} \rangle$~\cite{thooft78} can be used as a disorder 
parameter~\cite{vort1,vort2},
(order parameter of the disordered phase) corresponding to the dual
 $\tilde{Z}_N$ symmetry.

In full QCD, {\em i.e.} in the presence of dynamical quarks, the situation
is less clear: $Z_N$ and $\tilde{Z}_N$ symmetries are explicitely broken. 
At zero quark mass there is a 
phase transition at some $T_c$ where chiral symmetry is restored, the chiral
condensate being the order parameter: at the same $T_c$
also the susceptibility of the Polyakov loop 
shows a maximum which can be related to deconfinement.

However it is not clear 
theoretically what the chiral transition has to do with the
deconfinement transition. Moreover, in the presence of physical quark masses
also the chiral symmetry is explicitely broken, so that neither the 
Polyakov loop nor the chiral condensate are true order parameters.

In a series of papers \cite{artsu2,artsu3,artran}
we have tested and verified in the quenched case the mechanism of 
confinement based on dual
superconductivity of the QCD vacuum proposed by 't Hooft~\cite{thooft81}.
More precisely we have constructed
an operator $\mu$ 
which creates a magnetic monopole 
in a $U(1)$ subgroup of the color gauge group selected by abelian projection.
If the magnetic symmetry is realized
{\em \`a la} Wigner and if $\mu$ carries
non zero net magnetic charge, then $\langle \mu \rangle = 0$. 
Therefore $\langle \mu \rangle \neq 0$
implies Higgs breaking of the magnetic U(1) symmetry and thus
dual superconductivity.
It has indeed been shown that the quenched QCD vacuum
is a dual superconductor $\left( \pardis \neq 0 \right)$ in the confined phase
and 
goes to normal $\left( \pardis = 0 \right) $ at the deconfinement transition.
It has also been shown that $\pardis$ being different or equal to zero is 
independent
of the particular abelian projection chosen~\cite{artran,abelind}. 

Both the construction of the disorder parameter $\pardis$ 
and the related magnetic symmetry stay unaltered if
dynamical fermions are introduced: 
$\pardis$ is therefore a good candidate disorder parameter 
also in full QCD, especially if one expects, in the spirit
of the $N_c \to \infty$ limit, that
the mechanism which drives confinement and the
deconfinement phase transition be the same with and without
dynamical quarks.

In Ref.~\cite{artfull} we have indeed demonstrated that also in full QCD with
two light flavors, $\pardis$ shows a transition from a low temperature phase,
where it is different from zero thus signaling dual superconductivity,
to a high temperature phase where it is exactly zero, thus signaling the 
disappearence of dual superconductivity, or deconfinement.
We have also demonstrated that the transition for $\pardis$ coincides with the
chiral phase transition.
Analogous results have been obtained in Ref.~\cite{bari} by measuring  
a parameter related to the monopole free energy.

The aim of the present work is to extend the analysis and to show
that $\pardis$ scales with the correct
critical indices at the phase transition, thus demonstrating 
that it is indeed a valid disorder parameter.
We will discuss the results of a finite size scaling analysis of $\pardis$
around the phase transition, 
showing that it gives critical indices which are consistent with
those found through finite size scaling of the specific heat 
(and of the chiral susceptibility), as
presented in~\cite{lat04}.

\section{Disorder parameter}

The operator $\mu$ is defined in full QCD exactly in the same way as in the
quenched theory \cite{artsu2,artsu3,artran}
\beq
\label{defmu}
\langle \mu \rangle &=& \frac{\tilde{Z}}{Z} \; ,\nonumber \\
Z &=& \int \left( {\cal D}U \right)  e^{-\beta S} \; ,\nonumber \\   
\tilde{Z} &=& \int \left( {\cal D}U \right)  e^{-\beta \tilde{S}} \; .
\eeq
$\tilde{Z}$ is obtained from $Z$ by changing the action in 
the time slice $x_0$, $S \to \tilde{S} = S + \Delta S$.
In the Abelian projected gauge the plaquettes
\beq
\Pi_{i0} (\vec{x},x_0) = \nonumber 
\eeq
\beq
U_i (\vec{x},x_0) U_0 (x + \hat{\imath},x_0)
U_i^\dagger(\vec{x},x_0 + \hat{0}) U_0^\dagger (\vec{x},x_0)
\eeq
are changed by substituting
\beq
U_i(\vec{x},x_0) \to \tilde{U}_i(\vec{x},x_0) \equiv
U_i(\vec{x},x_0) e^{i T b_i (\vec{x} - \vec{y})}
\eeq
where $\vec{b}  (\vec{x} - \vec{y})$ is the vector potential of a 
monopole configuration centered at $\vec{y}$ in the gauge
$\vec{\nabla} \vec{b} = 0$, and $T$ is the diagonal 
gauge group generator corresponding to the monopole species chosen.
It can be shown that, as in the quenched case, 
$\mu$ adds to any configuration the monopole configuration
$\vec{b} (\vec{x} - \vec{y})$. 

Instead of $\pardis$ we measure the quantity
\beq
\rho = \frac{d}{d \beta} \ln \langle \mu \rangle \; .
\eeq
It follows from Eq. (\ref{defmu}) that
\beq
\rho = \langle S \rangle_S -  \langle \tilde{S} \rangle_{\tilde{S}} \; ,
\label{rhoferm}
\eeq
the subscript meaning the action by which the average is performed. 
In terms of $\rho$
\beq
\label{mufromrho}
\pardis = \exp\left(\int_0^{\beta} \rho(\beta^{\prime})\mbox{d}\beta^{\prime}\right) \; .
\eeq
A drop of $\pardis$ at the phase transition corresponds to a strong
negative peak of $\rho$.

\section{Numerical Results}
\label{sec:numres}

We have made simulations with two degenerate flavors of Kogut-Susskind quarks, 
using the standard gauge and fermion actions.
Configuration updating was performed using the standard Hybrid R algorithm. 
The lattice temporal size was fixed at $N_t=4$.
Different spatial sizes ($L = 12,16,20,24,32$) and 
values of the quark mass were used. 
For a more detailed account on simulation parameters we refer 
to~\cite{lat04}.

We can assume the following general scaling form for $\pardis$ around
the phase transition:
\beq
\pardis = L^k \Phi ( \tau L^{1/\nu},m L^{y_h} ) \nonumber
\eeq
where $\tau$ is the reduced temperature and $m$ the quark mass.
Analyticity arguments~\cite{lat04,kar} suggest that 
in the infinite volume limit the mass
dependence in the scaling function factorizes, so that 
$\rho = \frac{d}{d \beta} \ln \pardis$ 
does not depend on the mass.
We then obtain the following scaling law:\\
\beq
\rho = L^{1/\nu} \phi(\tau L^{1/\nu}) \; .
\eeq

In Figure 1 we show the quality of scaling assuming
$\nu = 1/3$, i.e. a first order phase transition: a good agreement
is clearly visible.
The deviation from scaling in the deconfined region is well
understood~\cite{artsup} and related to the disorder parameter
$\pardis$ being exactly zero on that side.

\begin{figure}[t!]
\includegraphics*[width=\columnwidth]{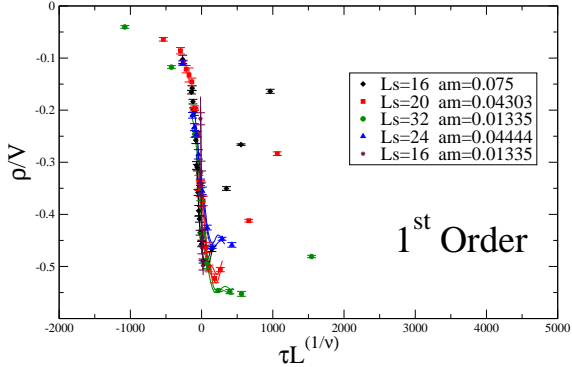}
\caption{Finite size scaling of $\rho$ according to first order 
critical indices.}\label{fig1st} 
\end{figure}

For comparison we show in Figure 2 the quality of scaling
assuming the $O(4)$ critical index $\nu = 0.75$.

The $O(4)$ universality class is clearly excluded, while there is
good agreement with a first order phase transition, confirming
results obtained through an analysis of the specific heat and 
of the chiral susceptibility~\cite{lat04}.

\section{Conclusions}

We have investigated the scaling properties of the parameter
$\pardis$, detecting dual superconductivity of the vacuum, around
the phase transition of full QCD with 2 flavors of staggered fermions,
in order to check if $\pardis$ behaves as a good disorder parameter
for the confinement - deconfinement phase transition, as it does in 
quenched QCD.

We have shown that a finite size scaling analysis of $\pardis$ indicates
a first order phase transition and excludes $O(4)$ critical behaviour,
in agreement with previous hints~\cite{artfull,bari} and with a 
detailed analysis of the specific
heat and the chiral susceptibility~\cite{lat04}.
This confirms that also in full QCD $\pardis$ can be considered 
as a valid disorder parameter.

\begin{figure}[t!]
\includegraphics*[width=\columnwidth]{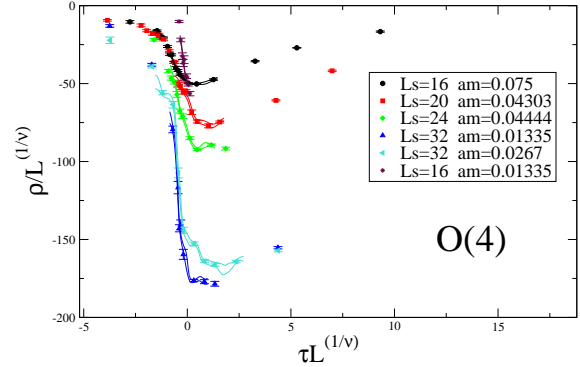}
\caption{Finite size scaling of $\rho$ according to $O(4)$ 
critical indices.}\label{figO4} 
\end{figure}

\end{document}